\documentclass[numberedappendix]{emulateapj}
\shorttitle{MORPHOLOGY AND SIZE OF LUMINOUS INFRARED GALAXIES}
\shortauthors{RUJOPAKARN ET AL.}

\newcommand{\Lsun}{\mbox{$L_\odot$}}
\newcommand{\LTIR}{\mbox{$L({\rm TIR})$}} 

\newcommand{\qir}{\mbox{$q_{24}$}} 
\newcommand{\SFRSD}{\mbox{$\Sigma_{\rm SFR}$}} 
\newcommand{\LIRSD}{\mbox{$\Sigma_{\rm L(TIR)}$}} 

\slugcomment{Accepted for publication in The Astrophysical Journal,
  September 28th, 2010} 
\begin{document}
\title{MORPHOLOGY AND SIZE DIFFERENCES BETWEEN LOCAL AND HIGH REDSHIFT
  LUMINOUS INFRARED GALAXIES} 

\author{WIPHU RUJOPAKARN\altaffilmark{1}, 
GEORGE H. RIEKE\altaffilmark{1}, 
DANIEL J. EISENSTEIN\altaffilmark{1}, 
ST{\'E}PHANIE JUNEAU\altaffilmark{1}}

\altaffiltext{1}{Steward Observatory, The University of Arizona,
  Tucson, AZ, USA 85721; wiphu@as.arizona.edu}

\begin{abstract}
We show that the star-forming regions in high-redshift luminous and
ultraluminous infrared galaxies (LIRGs and ULIRGs) and submillimeter
galaxies (SMGs) have similar physical scales to those in local normal
star-forming galaxies. To first order, their higher infrared (IR)
luminosities result from higher luminosity surface 
density. We also find a good correlation between the IR luminosity and
IR luminosity surface density in starburst galaxies across over five
orders of magnitude of IR luminosity from local normal galaxies to $z  
\sim 2$ SMGs. The intensely star-forming regions of local ULIRGs are
significantly smaller than those in their high-redshift counterparts
and hence diverge significantly from this correlation, indicating that
the ULIRGs found locally are a different population from the
high-redshift ULIRGs and SMGs. Based on this relationship, we suggest
that luminosity surface density should serve as a more accurate
indicator for the IR emitting environment, and hence the observable
properties, of star-forming galaxies than their IR luminosity. We
demonstrate this approach by showing that ULIRGs at $z \sim 1$ and a
lensed galaxy at $z \sim 2.5$ exhibit aromatic features agreeing with
local LIRGs that are an order of magnitude less luminous, but have
similar IR luminosity surface density. A consequence of this
relationship is that the aromatic emission strength in star-forming
galaxies will appear to increase at $z > 1$ for a given IR luminosity
compared to their local counterparts.
\end{abstract}
\keywords{galaxies: evolution --- galaxies: high-redshift ---
  infrared: galaxies}

\section{INTRODUCTION}
Luminous and Ultraluminous Infrared Galaxies (LIRGs and ULIRGs) whose
total infrared luminosities, \LTIR, are in the range of
$10^{11}-10^{12}$ and $>10^{12}$ \Lsun, respectively, are among the
most important populations for the study of galaxy evolution. Their
extreme \LTIR\ signifies intense star formation hidden by
dust and only visible through the infrared (IR) emission resulting from
the reprocessing of UV photons that originate from populations of
rapidly forming hot young stars. They are among the most extreme
star-forming regions known. 

A majority of the local ULIRGs are disturbed systems of galaxies
undergoing mergers. Some ULIRGs harbor active galactic nuclei (AGN)
often thought to be associated with merger activity and the resulting  
enhanced dense molecular gas fractions observed in the nuclei of these
systems \citep[e.g.][]{GS04,Juneau09}. Locally, luminous
IR galaxies are very rare \citep{Lagache05} but they contribute
significantly to the 
total IR energy density beyond redshift $z \sim 1$
\citep[e.g.,][]{LeFloch05, PPG05, Rodighiero10}. At high redshift, $z
\gtrsim 2$, 
luminous IR galaxies are typified by submillimeter galaxies 
 \citep[SMGs; e.g.][]{Blain02}. So far it is common to view SMGs as a
 more luminous counterpart of local ULIRGs. This naturally leads to
 the view of SMGs being disturbed, interacting systems
 \citep[e.g.][]{Tacconi08}. 

However, there are many indications that the SEDs of the high
redshift infrared galaxies differ systematically from those of local
galaxies of similar luminosity.  \citet{Papovich07} found that the 70
$\micron$ (observed) outputs tended to be weaker relative to those at 24
$\micron$ (observed) than expected from local templates. Many authors
have found that the aromatic bands in these galaxies appear to be
similar in structure to those in significantly lower-luminosity local
galaxies \citep[e.g.,][]{Rigby08, Farrah08, Takagi10}. The far
infrared SEDs appear to be cold, again similar to those of lower
luminosity local galaxies \citep[e.g.,][]{Pope06, Symeonidis09,
Muzzin10}. These findings suggest an underlying physical difference
between local luminous infrared galaxies and those at high 
redshift. \citet{Rigby08} suggested that such a  
difference might arise either through reduced metallicity or lower
optical depth due to a greater extent of the emitting 
regions. \citet{Erb06} find that the metallicities
are of order three times lower at $z \sim 2$ for a given galaxy
mass. \citet{Engelbracht08} show that, for local galaxies, there is
only a weak trend in 8 $\micron$ vs. \LTIR\ down to
$1/3$ solar metallicity, and that at lower metallicity the 8 $\micron$
luminosity is suppressed. This correlation is also reported for $z
\sim 2$ galaxies by \citet{Reddy10}. We conclude that
reduced metallicity is unlikely to be the primary cause of the changes
in SED with redshift. We therefore turn our attention to the second
possibility, that the galaxies have structures different from local
ones of similar luminosity.

Recent high-resolution studies of SMGs in the
submillimeter, radio, and near-IR have shown that their star-forming
regions are generally relatively extended, with diameters of order
$1-10$ kpc \citep{Chapman04, Muxlow05, BiggsIvison08, Bothwell09,
  Casey09, Iono09, Lehnert09, Carilli10, Swinbank10a, Tacconi10,
  Younger10}. Thus, the 
surface densities of the star formation rate (SFR) must be
substantially lower in the high-z galaxies than for the local ones of
similar \LTIR. This paper explores the importance 
of this difference in explaining the different SED behavior. We make
the simplest possible assumption, that the SED is a function of the
SFR surface density, \SFRSD, and that galaxies with similar \SFRSD,
and consequently similar \LTIR\ surface density, \LIRSD, will
have similar optical depths in their star forming regions and similar
SEDs. In section \ref{sec:data}, we describe the compilation of data
for the study as well as discuss our derivation of a consistent set of
sizes for the star forming regions for $0 < z < 2.5$. Section
\ref{sec:results} shows the results and discusses the implications for
both galaxy evolution and for improved estimates of the SFRs of high
redshift IR galaxies. We assume a $\Lambda$CDM cosmology with
$\Omega_m = 0.3$, $\Omega_{\Lambda} = 0.7$, and $H_0 =
70$~km~s$^{-1}$Mpc$^{-1}$ throughout this paper.

\section{THE MEASUREMENTS}\label{sec:data}
A major challenge in studying the \LIRSD\ of
galaxies across a broad redshift range is to obtain physical size
information for the IR-emitting regions in a consistent
way for both the local and high-redshift samples. We discuss
in this section the compilation of physical sizes and IR luminosity
data from the literature and our steps to compare them on the same
metric. Despite the abundantly available high-quality data for local
galaxies, it is necessary to select a subset that can
best match the high-redshift sample. We therefore will begin our
discussion with the high-redshift compilation, then follow with the
local compilation.

\subsection{Intermediate and High-Redshift
  Compilation}\label{sec:data_highz} 
The past five years have seen an unprecedented number of
high-resolution, multiwavelengths observations of high-redshift
galaxies. Our compilation of the physical sizes of intermediate and
high-redshift starburst galaxies is based on radio interferometric
observations at the Multi-Element Radio Linked Interferometer Network
(MERLIN) and the Very Large Array (VLA), as well as submillimeter
interferometric observations at the IRAM Plateau de Bure
Interferometer (PdBI). Submillimeter and 1.4 GHz observations, rather
than rest-frame optical or near-IR, are chosen because they closely
trace the star-forming regions of galaxies while being little affected
by the old stellar light. The 
submillimeter is a more direct tracer in this regard because it
observes thermal emission from dust while the 1.4 GHz data probe
synchrotron radiation from supernovae remnants, whose progenitors are
short-lived massive stars and hence indirectly trace
star-formation. Because of the radio-infrared relation within galaxies
\citep[e.g.,][]{Murphy06}, radio observations should give a
valid measure of the size of the active region in a star forming
galaxy. \citet{Ivison10a} report that the IR-radio relation could
evolve at high redshifts. However, their reported evolution law given
by $(1+z)^{-0.15 \pm  0.3}$ is small and unlikely to affect
significantly our estimation of IR-emitting region size using 
radio observations. \citet{Sargent10}, along with a recent result
using {\it Herschel} by \citet{Ivison10b}, also found little or no
evolution in the IR-radio relation at $z < 2$.

Although submillimeter transitions (e.g. CO) are more direct
tracers of molecular clouds and star-forming regions than is radio
emission,  they tend to trace colder gas. In comparison, the
star-forming regions in most local LIRGs and ULIRGs are sub-kpc in
size \citep[e.g.,][]{Condon91}. Thus the physical size given
by submillimeter observation tends to be systematically larger than
that of radio observations. Moreover, selection of high-redshift
submillimeter galaxies will tend to result in a sample with colder
SEDs, which are brighter in submillimeter wavelengths, and hence an
inherently physically extended sample. This bias towards larger
physical size should be more pronounced for the low$-J$ CO transitions
such as CO ($1-0$) and CO ($2-1$) and thus we will adopt sizes from
higher$-J$ CO transitions such as CO ($3-2$) and CO ($6-5$) if
available. To be conservative, we will take the radio sizes to be the 
physical size of the star-forming region in galaxies and treat
submillimeter sizes as upper limits to isolate this possible selection
effect. 

\subsubsection{High-z Submillimeter Compilation}\label{sec:data_hz_smm}
The submillimeter data were compiled from \citet{Tacconi06},
\citet{Tacconi10}, and \citet{Daddi10}. These authors used the IRAM
PdBI to study submillimeter-selected samples using CO transitions with
$0\farcs6$ to $\sim 1\farcs0$ resolution. \citet{Tacconi06} observed
six SMGs, including four sources from \citet{Greve03} and
\citet{Neri03}, and two new sources at redshifts $2.2 < z <
3.4$. Their observations yield four resolved SMGs with a median FWHM
diameter of $\sim 4$ 
kpc. One of the \citet{Tacconi06} sources needs to be excluded because
its radio emission is significantly stronger than predicted by
the IR-radio relation, indicating the presence of an AGN. \citet{Tacconi10}
reported the physical sizes (effective CO diameters) of two sources in
the Extended Groth Strip at redshifts $z \sim 1.1$ to be 13 and 16 kpc
in diameter, which they have found to be smaller but consistent with
the size based on an I-band observation. It should be noted that although
\citet{Tacconi06} find the SMGs to be ``compact'', they still
generally are a factor of 100 larger in area than
local ULIRGs of similar IR luminosities. Additionally, \citet{Daddi10}
reported sizes of four BzK-selected $z \sim 1.5$ 
star-forming galaxies (one member of this sample also has 1.4 GHz size
measurement from \citet{Casey09}, which we adopt in preference to the
CO data). In total we have eight data points from the submillimeter
observations; three from 
\citet{Tacconi06}, two from \citet{Tacconi10}, and three from
\citet{Daddi10}. IR luminosities for these eight galaxies were
estimated via the \citet{Rieke09} formalism based on 24 $\micron$
fluxes from \citet{Hainline09}, \citet{MD09}, and {\it Spitzer} deep
imaging in Great Observatories Origins Deep Survey (GOODS, Dickinson
et al., in prep.). 

\begin{figure}
\epsscale{1.22}
\figurenum{1}
\plotone{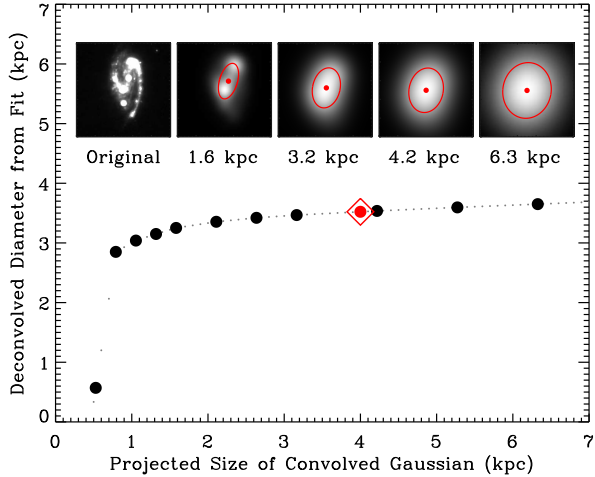}
\caption{To determine the physical size of star-forming regions
  consistently for both local and high-redshift galaxies, we need to
  degrade the image of local galaxies to resolution similar to
  that observed at high-redshift. We convolve local images with a 
  Gaussian beam of 4 kpc at the galaxy's distance and fit the
  convolved images to estimate the 
  deconvolved diameters. A typical case, NGC 3627, is shown here with
  the original {\it Spitzer} 24 $\micron$ image at the left panel of
  the inset. Subsequently we show the convolved images with the
  corresponding sizes of the convolved Gaussian beam and the 2D
  Gaussian fits to the convolved images overplotted as ellipses. This
  example demonstrates that the estimated deconvolved size is
  relatively insensitive to the choice of Gaussian used to convolve
  the original image.}
\label{local_conv}
\end{figure}

\subsubsection{High-z Radio Compilation}\label{sec:data_hz_radio}
The 1.4 GHz radio observations were drawn from \citet{Chapman04,
  Muxlow05, BiggsIvison08} and \citet{Casey09}. \citet{Chapman04} used
MERLIN+VLA with $0\farcs3$ resolution to study 12 SMGs selected from
the HDF at a median redshift of $z = 2.2 \pm 0.2$. $8/12$ of their
targets were resolved with an effective median diameter measured above
the $3\sigma$ contour of $7.0 \pm 1.0$ kpc. \citet{BiggsIvison08} used
MERLIN+VLA to observe 12 
SMGs in the Lockman Hole at $1.2 < z < 2.7$ with $0\farcs2-0\farcs5$
resolution (all sources were resolved). The sample was selected from
SMGs found by surveys carried out with SCUBA and MAMBO \citep{Scott02,
  Greve04, Coppin06}. \citet{BiggsIvison08} reported physical sizes of
these SMGs ranging from $1-8$ kpc with a median of 5 kpc. Lastly,
\citet{Casey09} used MERLIN ($0\farcs3-0\farcs5$ resolution) to observe
seven ULIRGs and a HyLIRG selected by optical and radio color criteria
from GOODS and the Lockman Hole at redshifts $0.9 < z <
2.4$. They resolved all eight and report an average physical diameter
of $5.0 \pm 1.2$ kpc, corresponding to the surface area within the
$3\sigma$ detection level. 

\citet{Muxlow05} used MERLIN+VLA to conduct a deep 1.4 GHz survey of
the HDF and Hubble Flanking Field (HFF) and studied 92 
radio sources at $0.18 < z < 4$ with $0\farcs2-0\farcs5$ resolution. All
but one source was resolved. For this study, we exclude sources that
\citet{Muxlow05} classified as AGN or AGN candidates. We also exclude
the Muxlow sources that are observed only by the VLA and hence have
lower resolution that may affect the accuracy of size estimates, as
well as those with complex morphology such that their size estimation
requires visual inspection \citep{Muxlow05, Thrall07}. Out of
92 objects from the \citet{Muxlow05} sample, 72 have spectroscopic
redshifts and 27 pass the aforementioned criterion and have a 24
$\micron$ counterpart in GOODS. \citet{Muxlow05} reported sizes
in terms of the largest angular size determined by Gaussian fitting
\citep{Thrall07}. Taking the largest angular size, which is the
size of the major axis, as the diameter directly would overestimate
the surface area; we assume that the sizes of the minor axes of these
galaxies are 0.6 of the major-axis (the largest angular FWHM) and
then calculate the surface areas and circularized diameters for the
\citet{Muxlow05} sample as reported in Table \ref{table_highz}. This
value is an average ratio of minor-to-major axes of SMGs from the
\citet{BiggsIvison08} sample. An average physical diameter for the
\citet{Muxlow05} sample is $4.5$ kpc. \LTIR\ for the
\citet{Chapman04}, the \citet{Muxlow05}, and the \citet{Casey09}
samples in the GOODS field were estimated via the \citet{Rieke09}
formalism and the {\it Spitzer} 24 $\micron$ imaging in
GOODS. \LTIR\ for the \citet{BiggsIvison08} and the \citet{Casey09}
samples in other fields are also estimated using the \citet{Rieke09}
formalism based on the 24 $\micron$ fluxes reported by
\citet{Ivison07} and \citet{Casey09}.

Apparently there are two approaches in reporting galaxy sizes from the
radio measurements: the deconvolved FWHM \citep{Muxlow05,
  BiggsIvison08} and the circularized diameter enclosing the surface
area of the $>3 \sigma$ radio detections 
\citep{Chapman04,Casey09}. Despite the methodological differences
between these two approaches, they agree very well for the objects
with overlapping observations, especially in the HDF where there are
six such galaxies. This agreement is expected, since the area detected
above $3 \sigma$ is likely to be consistent with that of half-max
power given the typical levels of signal to noise ratio in the
observations. We also note a very good agreement of sizes from 1.4 
GHz and CO ($3-2$) observations of a HDF object, SMMJ123707+6214SW,
where observations from \citet{Muxlow05} and \citet{Tacconi06} overlap.

Although galaxies drawn from the aforementioned samples are likely
dominated by star-forming activity, we independently confirm this by
inspecting whether their ratio of 850 $\micron$ and 1.4 GHz fluxes
follows the IR-radio relation for starburst galaxies given by
\citet{Rieke09}. Galaxies with radio flux significantly stronger than
predicted by the IR-radio relation likely harbor radio-loud
AGN. Indeed, we have found that the only two objects with 
F$_{850}/$F$_{1.4} \gg 2$ are the most compact objects in our
compilation (consistent with AGN-domination) and thus we exclude
these objects (SMMJ163650.43+405734.5 and
SMMJ105207.49+571904.0). It is worth noting that this test is
unnecessary for the \citet{Muxlow05} sample, where both 
radio spectral slope and radio morphology are already employed to 
identify AGN and AGN candidates.

In total our compilation has 48 intermediate and high-redshift
starburst galaxies with a median redshift of 1.0 and a median size of
5.1 kpc. Our primary sample comprises the 27 galaxies from
\citet{Muxlow05}, which should provide an unbiased and complete
sample, selected by radio flux alone. The results from the additional
21 galaxies are consistent with those from the \citet{Muxlow05}
observations. 

\subsection{Local Compilation}\label{sec:data_local}
For local galaxies, there is no homogeneous set of radio or
submillimeter images suitable for our needs. Therefore, we use two
additional measures of the star formation rate, Paschen-$\alpha$
images \citep{AH06} and images at 24 $\micron$ from a number of
sources. As with the radio, both measures are not strongly affected by
extinction and are good tracers of star forming activity
\citep[e.g.,][]{Murphy06, Calzetti10}. Although radio images of local
galaxies have somewhat larger extent than these other indicators, the
high surface brightness areas that would dominate the high-redshift
measurement have very similar morphologies and sizes \citep{Murphy06}.

\begin{figure*}
\epsscale{1.0}
\figurenum{2}
\plotone{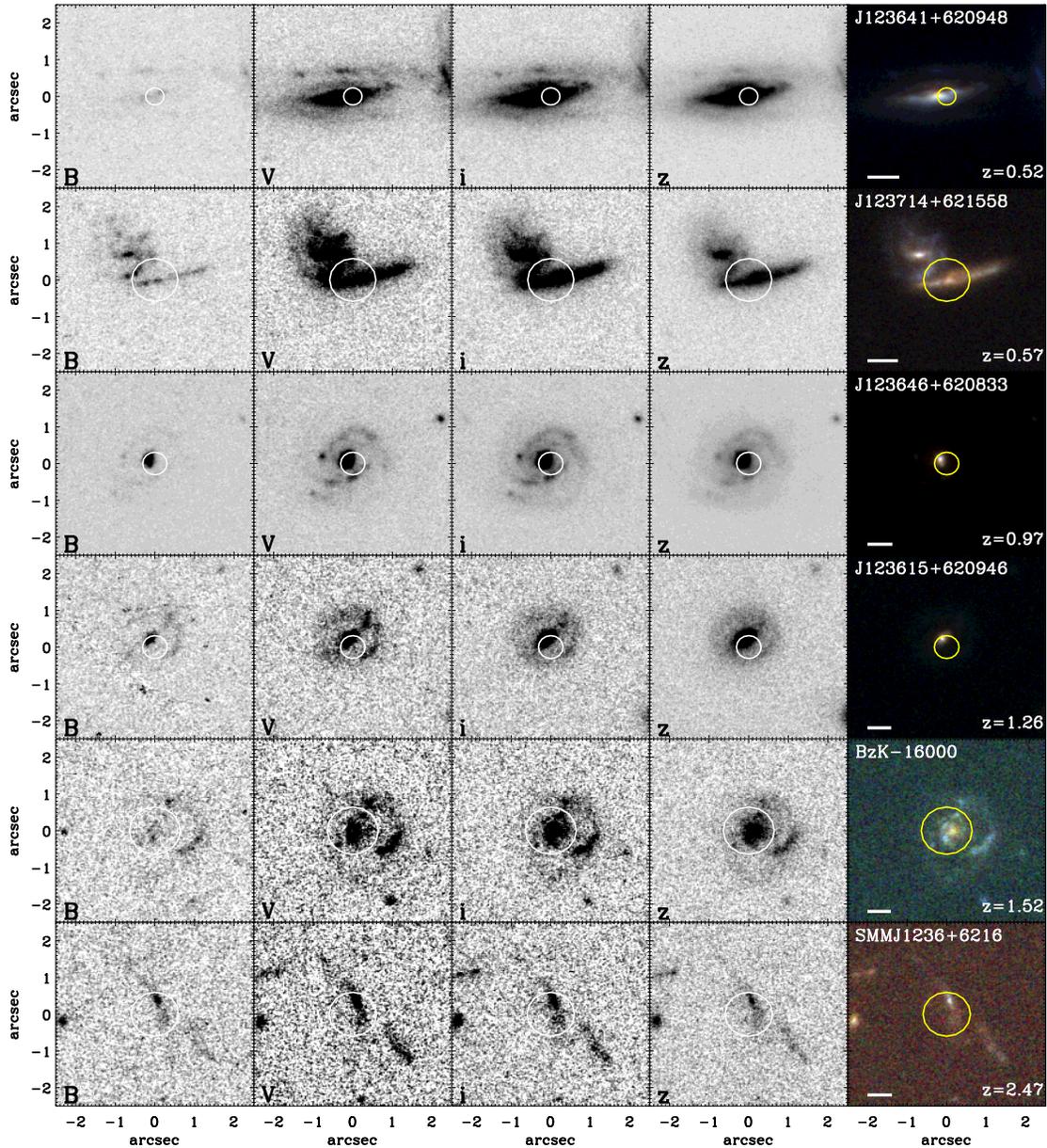}
\caption{{\it HST} ACS imaging of the subsample of our compilation in
  the GOODS field in B, V, $i$, and $z$. The circles indicate the size
  of the circularized diameter for the star-forming region in the
  corresponding galaxy. The bar at the lower left corner of the images
  in the right column represents a physical scale of 5 kpc. We found
  that $\sim2/3$ of the subsample in GOODS display isolated, quiescent
  star-forming galaxies while a few cases show obvious
  signs of galaxy interaction as well as disturbed morphologies that
  can be associated with asymmetric clumps of star formation,
  especially at high-redshift. [{\it See the electronic edition of the
      Journal for Fig. \ref{GOODS}.2$-$\ref{GOODS}.4.}]}
\label{GOODS}
\end{figure*}

The high redshift galaxies have limited structural information,
typically equivalent to a FWHM from Gaussian fitting or deconvolution
of the observed image, in which the galaxy is only modestly well
resolved. The FWHM of the beam is typically $0\farcs3$ to $1\farcs0$. A
$1\farcs0$ beam corresponds to a diameter of 5 kpc at $z = 0.4$, 7.5 kpc
at $z = 0.8$, and 8.5 kpc at $z = 2$. However, the images of local
galaxies often provide many more resolution elements across the
target. To put them on the same scale as the radio images at high
redshift, we convolved them with a Gaussian beam, and then determined
sizes by Gaussian fitting and deconvolution. We demonstrate with a
typical case of local star-forming galaxy in Fig. \ref{local_conv}
that this method yields a robust measurement of physical sizes for
local galaxies.
 
Despite our efforts to put all the images on the same basis, a
possible source of systematic error remains. An image with low signal
to noise will tend not to go as far into the wings of the star-forming 
activity and hence there will be a tendency for a systematic reduction
in the estimated FWHM with decreasing signal to noise. Thus, the
sizes of the high-redshift galaxies may be underestimated compared
with the local ones. Since the major result of this paper is that the
high-z galaxies have substantially larger sizes than local ones of
similar luminosity, the result is that we may understate this
conclusion, not that it would be undermined. 

We describe our
compilation of local IR galaxies in detail in the Appendix. It
comprises 19 normal star-forming galaxies ($\LTIR \leq 
10^{11}$ \Lsun), 21 luminous IR galaxies (LIRG, $10^{11} \leq \LTIR \leq
10^{12}$ \Lsun), and 4 ultraluminous IR galaxies (ULIRG, $\LTIR \geq
10^{12}$ \Lsun).

\subsection{{\it HST} Imaging of the High-Redshift Sample in the
  GOODS-N Field}\label{sec:data_GOODS} 
Since 36 of 48 high-redshift galaxies in our sample are in the {\rm
  Hubble} Deep Field imaged by the Great Observatories Origins Deep
Survey (GOODS; Dickinson et al., in prep.) using the Advanced Camera
for Survey (ACS) on the {\it Hubble} Space Telescope, we investigate the
GOODS imaging (Version 2.0) of these galaxies in
the B, V, $i$, and $z$ bands. Details about the ACS observations of
GOODS as well as the data can be found in \citet{Giavalisco04} and at
the survey's website\footnote[1]{http://www.stsci.edu/science/goods}.

GOODS optical images are used to confirm independently that the sizes
and positions of the star-forming regions determined by radio or
submillimeter observations are consistent with the high-resolution
optical observations. The synthesized beam sizes of the radio and
submillimeter observations are $0\farcs2 - 0\farcs5$
while the optical imaging in GOODS/ACS has resolution at
$\sim 0\farcs05/$pixel. Although the radio and submillimeter
observations can constrain the size of the star formation regions
without being affected by stellar emission, their interpretation
benefits from high-resolution optical imaging, especially for
morphological classifications. We illustrate the circularized size for
the star-forming regions compared to the optical 
extents of their host galaxies as seen in various optical filters in
Fig. \ref{GOODS}. The star-forming regions seen at
radio and submillimeter wavelengths typically coincide with the
central part of the optical structure, but there are a few cases where
the star-forming region is at the collision interface of an interacting
system (e.g. J123714+621558, J123716+621643). We will discuss our
qualitative assessment of optical morphologies to address the
structure of high-redshift star-forming galaxies in
\S~\ref{sec:results_unified}.

\begin{figure*}
\epsscale{0.8}
\figurenum{3}
\plotone{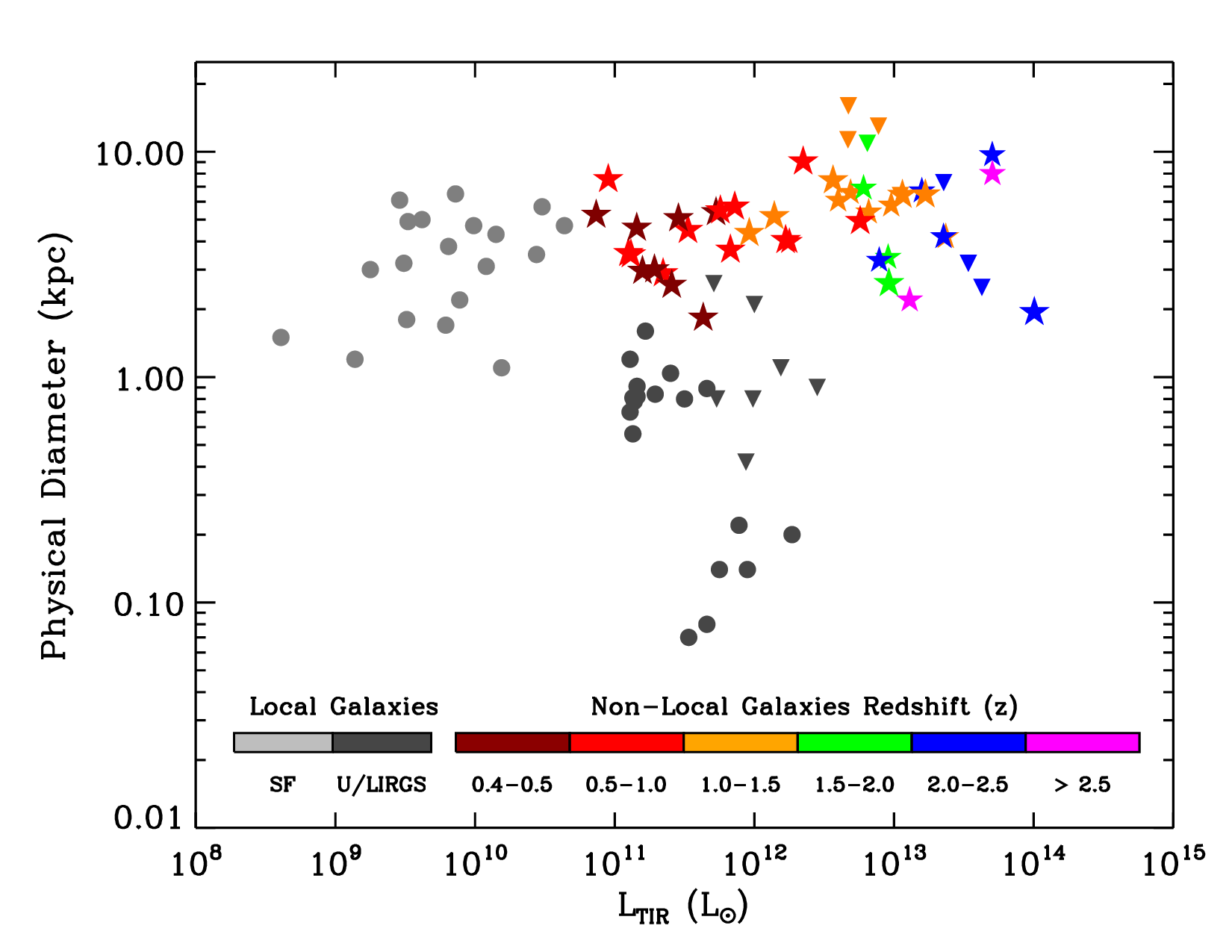}
\caption{Physical sizes of star-forming galaxies at various redshifts
  and IR luminosities. Galaxies shown in color-coded stars and grey
  dots have diameters determined by radio observations (24
  $\micron$ for the local normal star-forming galaxies), while those
  shown in color-coded downward triangles and grey downward triangles
  have diameters determined with CO observations, which could yield
  systematically larger values (particularly for low$-J$ CO transitions),
  the symbols signify upper limits in size. High-redshift LIRGs,
  ULIRGs, and SMG have similar sizes to local normal
  star-forming galaxies, while 
  local LIRGs and ULIRGs are significantly smaller.}  
\label{LTIR_Size}

\figurenum{4}
\plotone{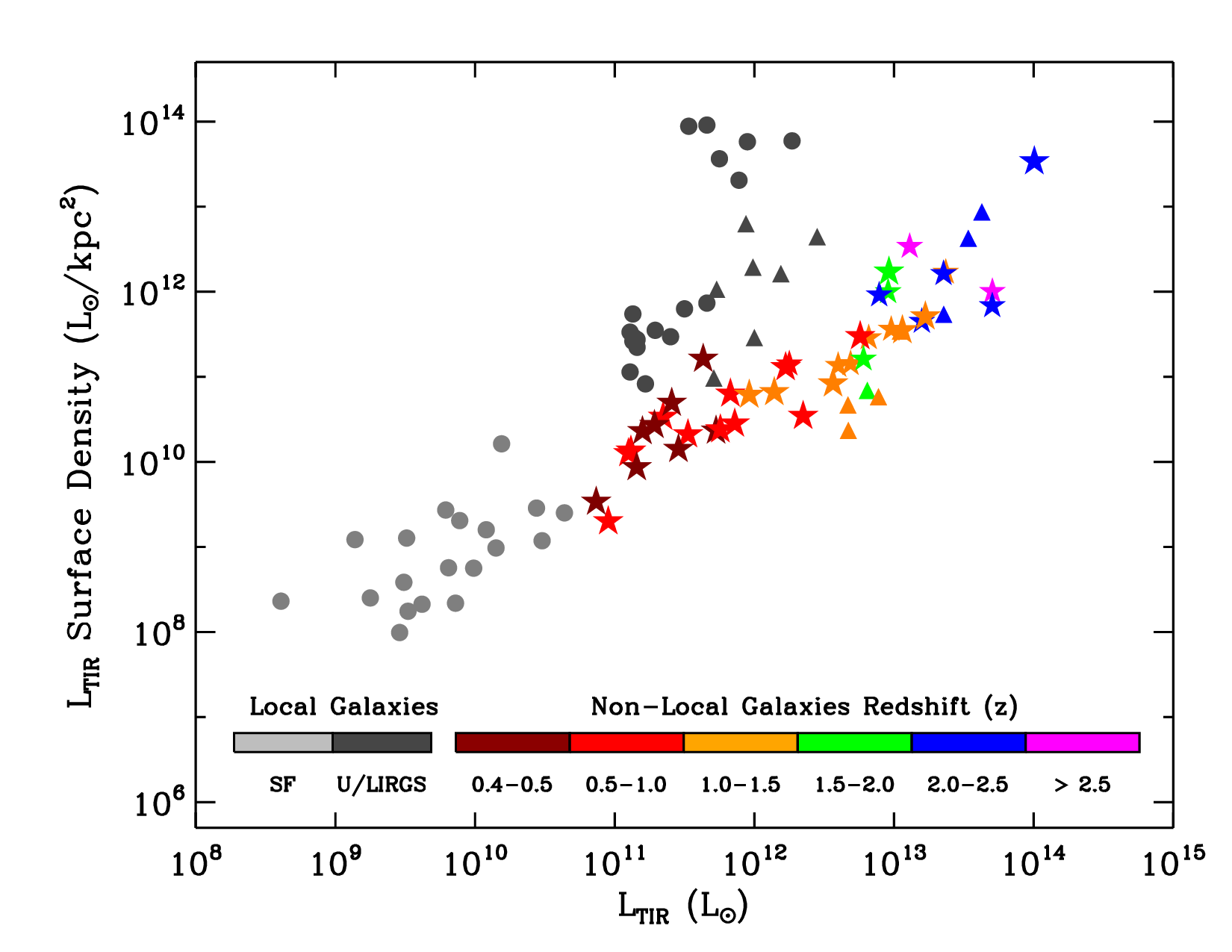}
\caption{IR luminosity surface density, \LIRSD, as a function of IR
  luminosity for star-forming galaxies at various redshifts with the same
  color scheme as Fig. \ref{LTIR_Size}. The correlation
  seen in local starbursts and intermediate- and high-redshift LIRGs,
  ULIRGs, 
  and SMGs indicates a general relationship in their conditions for
  star formation. On the other hand, local LIRGs and ULIRGs represent a
  different class of objects with star formation likely driven by
  some other process, such as galaxy interactions and mergers.}
\label{LTIR_density_z}
\end{figure*}

\section{RESULTS}\label{sec:results}
Our determinations of the physical sizes and luminosities for
star-forming galaxies from the local sample out to the high-redshift
SMGs are summarized in Fig. \ref{LTIR_Size}. We find the physical
sizes of star-forming galaxies to be comparable within an order of
magnitude across the entire IR luminosity ($10^9 - 10^{14}$ \Lsun) and
redshift range ($0 < z < 2.5$). Local LIRGs and ULIRGs, however, are
significantly smaller in size. It follows from Fig. \ref{LTIR_Size}
and is apparent from the diagram of \LIRSD\ as a function of \LTIR\ in
Fig. \ref{LTIR_density_z} that there is a correlation extending more 
than five orders of magnitude between \LIRSD\ and \LTIR. A linear fit
to this correlation yields a formal slope of $0.9$. However, given the
inhomogeneity of the methods employed to estimate uncertainties in
each of the subsamples comprising our compilation, we expect the
uncertainty for this formal slope to encompass the slope of unity and
hence suggest that the two parameters are proportional to each other.

Local LIRGs and ULIRGs have $1-4$ orders of magnitude higher \LIRSD\ than
indicated by this correlation suggesting that the starburst regions
in these galaxies are not representative of their high-redshift
counterparts of similar \LTIR.  In fact, local ULIRGs with
\LTIR\ $\sim 10^{12}$ have a comparable \LIRSD\ to high-redshift SMGs
with \LTIR\ $\sim 10^{14}$ \Lsun.

\begin{figure}
\epsscale{2.3}
\figurenum{5}
\plottwo{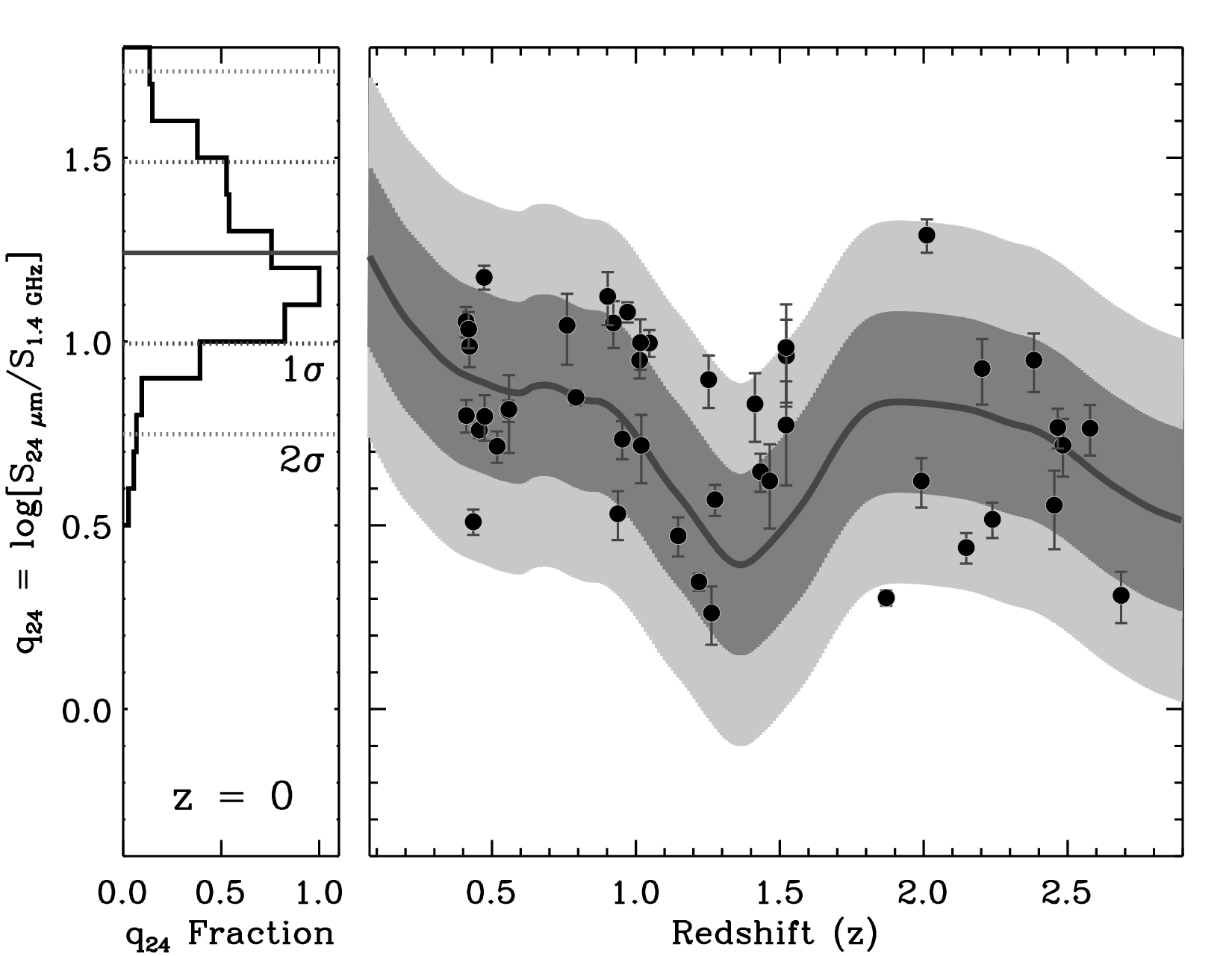}{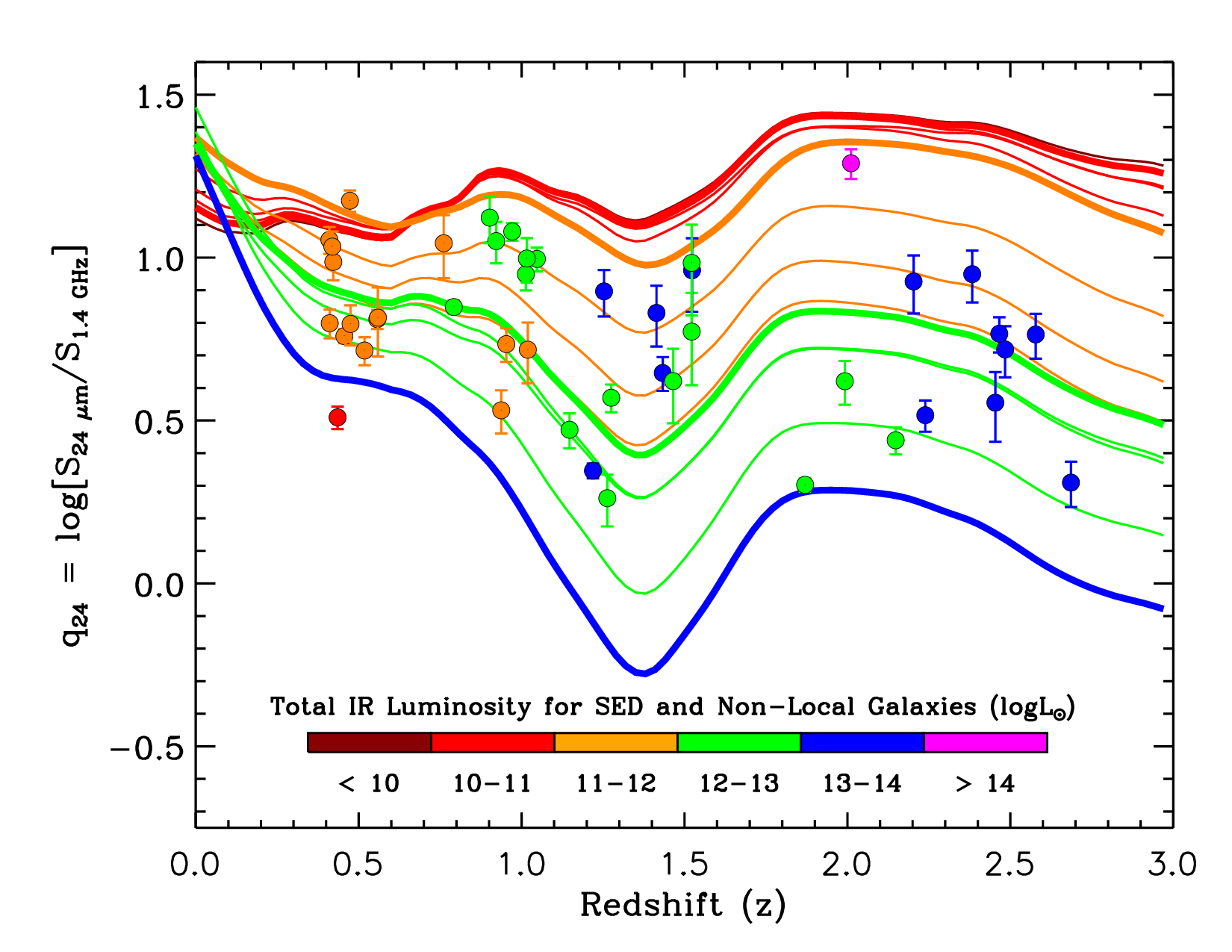}
\caption{(Top) The logarithmic ratio of 24 $\micron$ and 1.4 GHz flux
  densities, \qir, for  high-redshift star-forming galaxies ({\it
    right panel, black dots}) 
  with the local distribution of \qir\ ({\it left panel}). The local
  distribution is from the {\it IRAS} Bright Galaxy Catalog
  \citep{Sanders03} and the NRAO VLA survey \citep{Condon98}. The thick
  central line in the {\it right panel} shows the track of \qir\ predicted
  by the \citet{Rieke09} SED for a galaxy with \LTIR\ of $10^{12}$
  \Lsun, with dark and light grey shades showing the 1$\sigma$ and
  2$\sigma$ extents assuming the local distribution of \qir. Despite a
  relatively large scatter, the distribution of \qir\ at intermediate
  and high redshifts broadly agrees with the scatter observed
  locally. (Bottom) The logarithmic ratio of 24 $\micron$ flux and 1.4
  GHz flux predicted by the \citet{Rieke09} local LIRG and ULIRG SED
  templates ({\it solid lines}) compared to the observed ratios at
  $0.5 \lesssim z 
  \lesssim 2.5$. The observed ratios are consistent with the
  predictions from local templates for galaxies with significantly
  lower IR luminosities. For instance, galaxies with \LTIR\ 
  of $10^{13}-10^{14}$ \Lsun\ ({\it blue dots}) at $z \sim 2.5$ have a ratio
  consistent with the template for \LTIR\ $~\sim 10^{12}$ \Lsun\ ({\it
    green line}).}
\label{q24}
\end{figure}

\subsection{\SFRSD\ as a Tracer of Star-Forming
  Environment}\label{sec:results_implc}  
The aforementioned correlation suggests an explanation for the strong
aromatic emissions as well as the SEDs of star-forming galaxies at high
redshift. If we assume that \SFRSD, and 
hence \LIRSD, is an accurate indicator of the star-forming environment in
starburst galaxies, Fig. \ref{LTIR_density_z} would suggest that $z \sim
1$ ULIRGs have similar environments to local LIRGs with an order of
magnitude lower \LTIR, and likewise $z \sim 2$ SMGs have similar
environments to local ULIRGs with two orders of magnitude lower
\LTIR. We test this possibility with three independent approaches. 

First, we compare the logarithmic ratio of $24~\micron$ flux and 1.4
GHz flux, log(S$_{24}$/S$_{1.4}$) or \qir, predicted by the SEDs of
local LIRGs and ULIRGs and the observed flux ratio for high-redshift
galaxies. At redshift 
$1 < z < 2.5$ the {\it Spitzer} 24 $\micron$ observed bandpass probes
rest-frame wavelengths of $7-12~\micron$ and thus \qir\
at these redshifts is highly sensitive to emission from aromatic
complexes, which are sensitive to the star formation environment. We first
investigate the intrinsic scatter of \qir\ by considering the
scatter observed locally derived from the {\it IRAS} and VLA
observations, shown in the top panel of Fig. \ref{q24}. If we assume
a similar scatter around the mean value of \qir\ and project this
scatter to higher redshifts, the bottom panel of Fig. \ref{q24}
demonstrates that the observed scatter at high redshift is within the
expected scatter based on local observations. 

\begin{figure}
\epsscale{1.2}
\figurenum{6}
\plotone{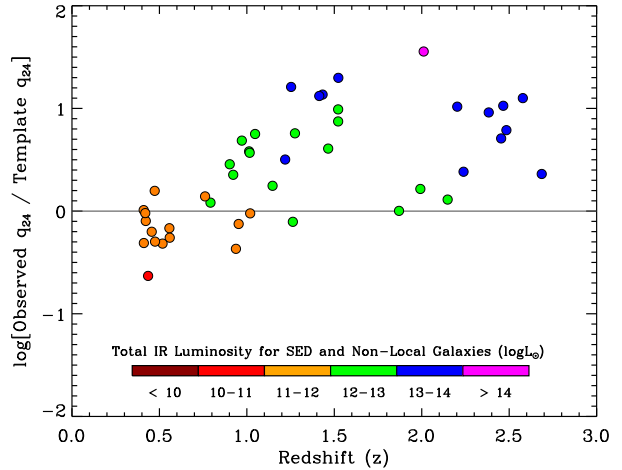}
\caption{The logarithmic ratio of the observed 24 $\micron$ and 1.4
  GHz flux ratios and those predicted by the \citet{Rieke09} SED
  templates based on local star-forming galaxies. Color coding
  represents the \LTIR\ for each object as in Fig. \ref{q24}. The
  increase of the 
  ratios above $z \sim 1$ indicates that the aromatic emissions at this
  redshift range are stronger than expected based on local SED
  templates.}
\label{q24_z_ratio}
\end{figure}

We then show in the right panel of Fig. \ref{q24} that \qir\ for
high-redshift galaxies is consistent with the ratios from local LIRGs
and ULIRGs 
with significantly lower \LTIR. Consider the SMGs at $z \sim
2.4$ with \LTIR\ $\sim 10^{13.5}$ \Lsun; the correlation in
Fig. \ref{LTIR_density_z} suggests their starburst environment to be
similar to local ULIRGs with \LTIR\ $\sim 10^{12}$ \Lsun\ and
\qir\ is clearly inconsistent with the extrapolation for local ULIRGs at
\LTIR\ $\sim 10^{13.5}$ \Lsun\ while agreeing with the SED for
\LTIR\ $\sim 10^{12}$ \Lsun\ local ULIRGs. Another way to interpret
this is illustrated by Fig. \ref{q24_z_ratio} that the aromatic
features are significantly stronger compared to local SED templates at
redshifts greater than $z \sim 1$.

Second, we consider a result from \citet{Rigby08}, specifically their
Fig. 3, that shows aromatic emissions for SMM J1635554.2+661225, a
lensed SMG at $z = 2.516$. This SMG has \LTIR\ of $10^{11.9}$ \Lsun,
which would be hardly detectable if not for the $22\times$
lensing magnification. The magnification also allows high
signal-to-noise mid-IR spectroscopy with {\it Spitzer}, in which
\citet{Rigby08} found aromatic emission band shapes similar to those
of NGC 2798 and to the average mid-IR spectra of 13 local normal
starburst galaxies with average \LTIR\ of $10^{10.7}$
\Lsun\ \citep{Brandl06,Dale07}. This result is consistent with the
relationship given in Fig. \ref{LTIR_density_z}; high-redshift
galaxies with \LTIR\ of $\sim 10^{11.9}$ \Lsun\ would have
\LIRSD\ similar to local galaxies with an order of magnitude lower
\LTIR. 

\begin{figure}
\epsscale{1.2}
\figurenum{7}
\plotone{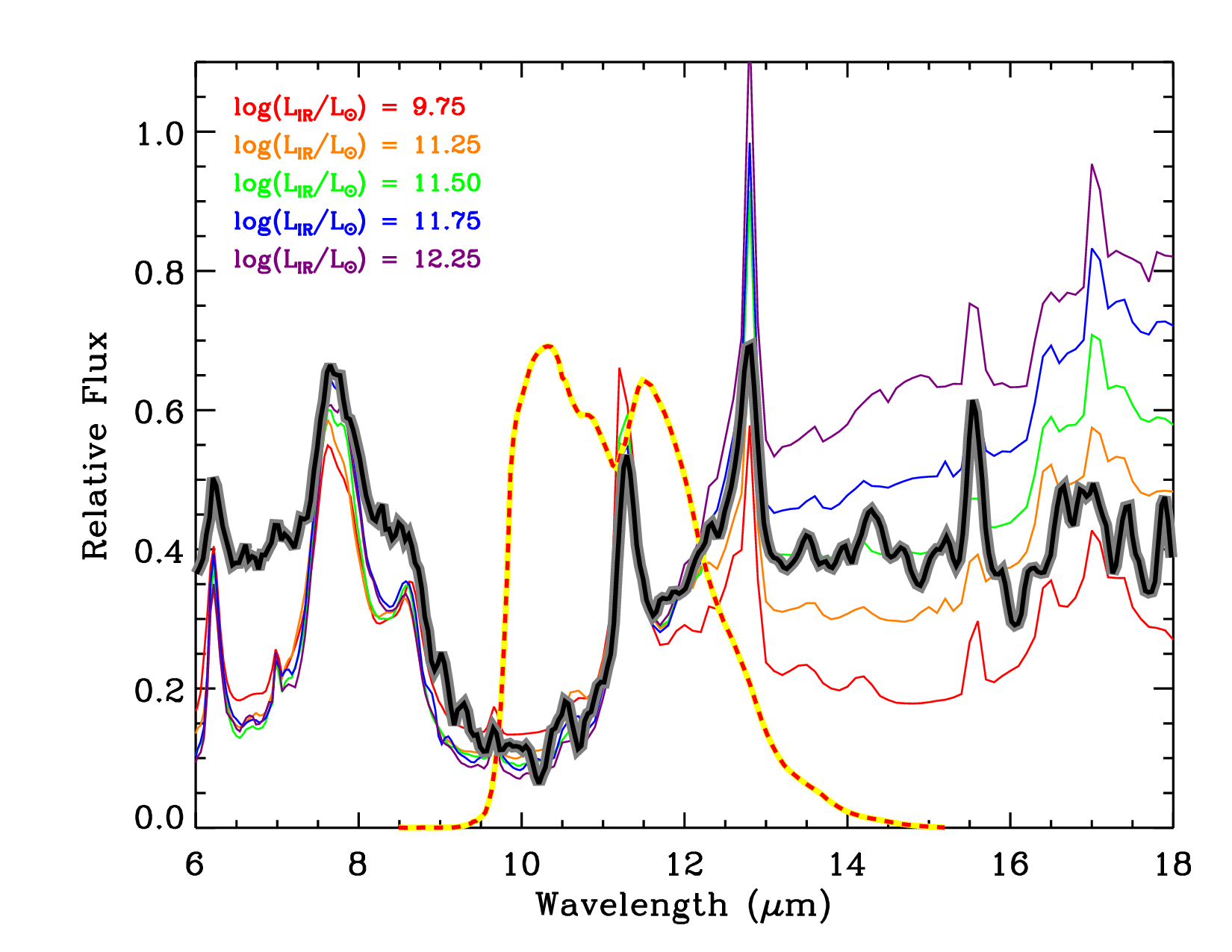}
\caption{The stacked observed spectrum of ULIRGs at $0.3 < z < 3.5$
  with average redshift $z = 1.1$ ({\it thick black line}) from
  \citet{Dasyra09} compared to the SED templates of \citet{Rieke09}
  derived for local galaxies ({\it color coded lines}). This plot is
  normalized by the flux at the observed 24 $\micron$ bandpass of {\it
    Spitzer} at $z = 1.1$ ({\it red dotted line}). The $z \sim 1$ 
  stacked ULIRG spectrum with \LTIR\ $\sim 10^{12.3}$ \Lsun\ has
  spectral features consistent with local galaxies that have 
  \LTIR\ $\sim 10^{11.3}$ \Lsun, agreeing with the prediction from the
  relationship in Fig. \ref{LTIR_density_z}.}
\label{Dasyra_smg}
\end{figure}

Third, we compare a stacked mid-IR spectrum of $z \sim 1.1$ ULIRGs
observed by \citet{Dasyra09} with the local SEDs from
\citet{Rieke09}. Fig. \ref{LTIR_density_z} suggests that the spectral
features of local \LTIR $\sim 10^{11.3}$ \Lsun\ galaxies should be
similar to those for $z \sim 1$ ULIRGs with \LTIR\ $\sim 10^{12.3}$
\Lsun. We confirm this prediction in Fig. \ref{Dasyra_smg}, where the
average observed spectrum is indeed consistent with local LIRGs with
\LTIR\ of $\sim 10^{11}$ \Lsun\ and clearly departs from the local SED
for $\sim 10^{12.25}$ \Lsun\ ULIRGs. 

These tests support our hypothesis that \LIRSD\ is a valid tracer
for the starburst environment and the observable spectral
features. More importantly, they independently confirm that
high-redshift star-forming galaxies, including SMGs, have similar
star-forming environments to local normal star-forming galaxies and
their higher star-formation rate is primarily due to higher
$\Sigma_{\rm SFR}$. 

\subsection{A Unified View of Star-Forming
  Galaxies}\label{sec:results_unified}
We believe from the nearly consistent sizes of local and high-redshift
star-forming galaxies that the most significant evolution between these
two populations is in the \SFRSD. In other words, while
the sizes of these galaxies remain about the same, their star
formation rate densities increase greatly from normal star-forming
galaxies to SMGs. 

Apart from many high-resolution observations of
high-redshift SMGs that find them to be physically extended systems,
recent observations of the 158 $\micron$ [CII] line provide another
probe into the environment of the star-forming regions. The [CII] line is
an important cooling line for the photodissociation regions at the 
surfaces of molecular clouds. Combining this [CII] line with the CO
($1-0$) line yields a color-color diagram of L$_{\rm [CII]}/$L$_{\rm
  IR}$ vs. L$_{\rm CO(1-0)}/$L$_{\rm IR}$ that 
is sensitive to both the incident UV flux and the gas
density. \citet{HD10} study the environment of the $z = 1.3$ galaxy
MIPS 1428 that has \LTIR\ $\sim 10^{13}$ \Lsun\ and report that its
L$_{\rm [CII]}/$L$_{\rm IR}$ is a factor of $\sim4$ higher than for
local ULIRGs while the L$_{\rm CO(1-0)}/$L$_{\rm IR}$ ratios are
comparable, indicating a similar incident UV flux in both populations
but that the gas density of MIPS 1428 is $\sim100 \times$ lower than
those in local ULIRGs. This behavior suggests a galaxy-wide
starburst. \citet{Ivison10c} apply this analysis using {\it Herschel}
measurements, resulting in a similar 
finding for SMMJ2135 at $z = 2.3$. The star-forming environments in
these two high-redshift galaxies appear to be similar to M82 and other
normal starburst galaxies rather than to local ULIRGs \citep{HD10,
  Ivison10c}. SMMJ2135 is a particularly noteworthy case because
it is lensed by $32\times$, which allowed \citet{Swinbank10b} to
observe it with the SMA at $0\farcs3 \times 0\farcs2$ resolution. They
resolved the galaxy into four $\sim100-$pc massive star-forming regions
distributed across the projected distance of 1.5 kpc, directly
confirming the distributed nature of the galaxy-wide star formation.

The optical morphologies of the subsample observed by GOODS, shown in
Fig. \ref{GOODS}, suggest that $\sim 2/3$ of the subsample are
quiescent, normal galaxies and the other $\sim 1/3$ show signs of
disturbed morphologies. The fraction of disturbed morphologies increases
with redshift. However, only $\sim 5$ systems out of 36 can be
identified positively as interacting systems, while the rest of those
with disturbed morphologies could as well be due to instabilities
fueled by rapid, asymmetric infall of gas resulting in large clumps of
star-forming regions similar to those seen in
SMMJ2135. \citet{Lehnert09} also observed clumpy, galaxy-wide
starbursts in 11 star-forming galaxies at $z \sim 2$ in rest-frame
optical wavelengths. The optical surface brightnesses for their sample
are more than an order magnitude greater than for local star-forming
galaxies, consistent with our result.

The molecular gas and star formation relation, the Kennicutt-Schmidt
Law \citep[e.g.][]{Schmidt59,Kennicutt98}, is shown by
\citet{Genzel10} to have a slope of 1.1 to 1.2 over a large range of
stellar mass surface density ($10^{0.5}$ to $10^4$
M$_{\odot}$pc$^{-2}$) for both low and high-redshift samples. A
remarkable difference between low and high-redshift star-forming
galaxies is that the gas depletion time increased from 0.5 Gyr at $z
\sim 2$ to 1.5 Gyr locally \citep{Genzel10}, which is consistent with
the picture that star-forming galaxies at low and high-redshift harbor
similar star-forming environments but the gas consumption rate, and hence
the star formation rate, is significantly higher at
high-redshift.

The finding that the physical conditions in high-redshift galaxies'
star-forming regions are 
similar to those in local quiescent star-forming galaxies indicates that
their intense star formation is unlike the transient starbursting
phase due to rapid infall of gas as a result of galaxy interaction, as
seen in local ULIRGs. Rather they may represent an isolated evolution
which could be observable for an extended period of time. This picture
is supported by the behavior of massive star-forming galaxies at
high-redshift found in the cosmological simulations of
\citet{Agertz09} and \citet{Dave10}. The latter simulated populations
with observational properties consistent with SMGs consists of
isolated galaxies in the middle of large potential wells with large
gas reservoirs. It should also be noted that \citet{Dave10} report a
highly asymmetric distribution of gas density, star formation, and
velocity field in the simulated SMGs consistent with the disturbed
morphologies observed.

\section{CONCLUSION}\label{sec:conclusion}
We made a compilation of physical size measurements for 43 local
galaxies with \LTIR\ ranging from normal star-forming
galaxies at $10^9$ \Lsun\ to ULIRGs at $> 10^{12}$ \Lsun, as well as
48 intermediate and high-redshift galaxies, including SMGs. 

Our compilation shows that (1) the physical scale of high-redshift
ULIRGs and submillimeter galaxies is consistent within an order of
magnitude with that of local normal star-forming galaxies (4.4 kpc in
median diameter), while local LIRGs and ULIRGs are significantly
smaller (0.8 kpc in median diameter); (2) there is a correlation of
\LTIR\ and \LIRSD\ extending over five orders of magnitude in
\LTIR\ for normal star-forming galaxies and high-z galaxies. Local
LIRGs and ULIRGs have significantly higher \LIRSD\ than high-redshift
galaxies with similar \LTIR\ and diverge from this correlation.

The fact that we do not find a significant deviation from this
relationship in high-redshift galaxies with \LTIR\ in the ULIRG range
indicates that the local ULIRGs as well as LIRGs belong to a rare
population likely driven by a unique process. High-resolution studies
of local ULIRGs have pointed  
out their disturbed morphology, double nuclei, and other signs of
merger activity. The correlation we have found, however, suggests that
the high \LTIR\ of SMGs and ULIRGs at large redshifts can be
explained, to first order, by the higher \SFRSD\ within isolated,
quiescent galaxies. 

We thank Anita Richards and Tom Muxlow for radio data in the HDF and
insightful discussions; Scott Chapman for radio images from
\citet{Chapman04}; Kalliopi Dasyra for spectra from \citet{Dasyra09};
Kar{\'i}n Men{\'e}ndez-Delmestre for spectra published in
\citet{MD09}. We also thank Benjamin Weiner for his assistance with
data compilation and valuable discussions. This work is supported by
contract 1255094 from Caltech/JPL to the University of Arizona. WR
gratefully acknowledges the support from the Thai Government Scholarship.

\appendix
\section{Estimation of Physical Sizes for Local IR Galaxies}
Our local sample of star-forming galaxies was selected to represent a
broad range of IR luminosities from normal star-forming galaxy ($\LTIR
\leq 10^{11}$ \Lsun), luminous IR galaxy (LIRG, $10^{11} \leq 
\LTIR \leq 10^{12}$ \Lsun), to ultraluminous IR galaxy (ULIRG,
$\LTIR \geq 10^{12}$ \Lsun). Here we discuss how we select the local
starburst galaxies and the methods we employ to estimate the physical
sizes to compare with high-redshift 
star-forming galaxies in the same metric. For all but three local
galaxies, \LTIR\ was obtained from the {\it IRAS} Revised Bright
Galaxy Sample \citep[RBGS;][]{Sanders03}. The RBGS provides \LTIR\ that
has a definition similar to that of \LTIR\ from the \citet{Rieke09}
formalism used for our intermediate and high redshift compilation. The
three exceptions are discussed separately.

\subsection{Normal Star-Forming Galaxies}\label{sec:data_local_NSF}
Galaxies from the {\it Spitzer} Infrared Nearby Galaxies Survey
\citep[e.g.,][]{Kennicutt03, Calzetti07} are used to represent
normal star-forming galaxies. SINGS galaxies are selected from within
the local volume to allow IR imaging at reasonably good physical
resolution. We use the 24 $\micron$ images of these galaxies to map
star formation. We only select high-luminosity galaxies. As noted by
\citet{Calzetti07}, NGC 4125 and NGC 5195 contain Seyfert 2 nuclei and
thus are excluded from our compilation. \LTIR\ for the other galaxies was
taken from the RBGS except for NGC 1512, NGC 2841, and NGC 4625 that
are not in the catalog. For these three galaxies we use 24 $\micron$
fluxes from \citet{Dale07} to estimate \LTIR\ using the appropriate
\citet{Rieke09} SED. 

We convolved the SINGS 24 $\micron$ MIPS images with a
series of Gaussians that have angular FWHM corresponding to physical
sizes ranging over $1 - 8$ kpc at the galaxy's distance and then fit 2D
Gaussians to estimate a deconvolved size from the convolved images. We
found that the recovered FWHM is a slowly varying function of the
convolved Gaussian's FWHM, which assures that sizes estimated from this
method are robust. For the actual size measurement for these galaxies
we convolve all SINGS 24 $\micron$ imaging with a Gaussian that has
FWHM corresponding to 4 kpc at each galaxy's distance and measure
deconvolved sizes from the convolved image. This procedure is
illustrated by NGC 3627 in Fig. \ref{local_conv}. These sizes are
reported in Table \ref{table_local} for normal star-forming galaxies.

We have 19 galaxies from the SINGS sample with \LTIR\ in the
range of $10^{8.6}-10^{10.6}$ \Lsun, a median \LTIR\ of $10^{9.8}$
\Lsun, and a median physical diameter of 3.5 kpc.

\subsection{LIRGs}\label{sec:data_local_LIRGs}
Our local LIRGs were drawn from \citet{Condon91}, \citet{AH06},
and \citet{Iono09}. Their size measurements are based on the 8.4 GHz
VLA radio observations, Pa$\alpha$ imaging using {\it Hubble} Space 
 Telescope (HST), and Submillimeter Array (SMA) submillimeter
 observations, respectively. 

\citet{Condon91} observed 40 LIRGs selected from the {\it IRAS} Bright
Galaxy Sample using the VLA with a resolution of $0\farcs25$. From
these 40 LIRGs, 15 are dominated by a compact radio component likely
to be an AGN, referred to by \citet{Condon91} as ``monsters'' and
another 5 are known to harbor AGN according to SIMBAD, which we
exclude; 6 are multi-component with one or more components not
reported, which we exclude; 2 are multi-component with well 
constrained sizes for both components (NGC 3690, which is Arp 299 for
which we adopt CO size measurement from \citet{Iono09}, and IRAS
F15163+4255); one has an observational issue (NGC 1614) but it was also
observed by \citet{Iono09} and therefore we adopt the
\citet{Iono09} size measurement for NGC 1614; there are size
measurements for only four galaxies of the remaining 11. These four
are UGC 2369, IRAS F03359+1523, UGC 4881, and IRAS F17132+5313. Along
with the aforementioned IRAS F15163+4255, a multi-component object with
well-measured size, we have five LIRGs from \citet{Condon91}.

The \citet{AH06} LIRGs were selected such that their Pa$\alpha$
emission line would fall in the narrow band of the F190N filter of the
Near Infrared Camera and Multi-Object Spectrometer (NICMOS) on {\it
  HST}. The resolution of NICMOS ($0\farcs076/$pixel) resolved
detailed structures of these LIRGs and thus we need to measure their
sizes using the same procedure as in
\S~\ref{sec:data_local_NSF}. However, we convolved the images using
Gaussians with FWHMs corresponding to only 1 kpc at each LIRG's
distance. This was necessary because of the intrinsically smaller
sizes of the LIRGs, which we also tested by measuring
the diameter encircling 90\% of flux. The results from our measurements
are given in the LIRGs section of Table \ref{table_local}.

\citet{Iono09} report sizes of LIRGs within 200 Mpc using the SMA, based on CO ($3-2$) observations by \citet{Wilson08}. The
sample contains LIRGs harboring AGN (including NGC 2623, NGC 6240, UGC
5101, Mrk 231, Mrk 273, IRAS F10565+2448 and IRAS F17207-0014) and
multicomponent objects where the sizes of some components cannot be
determined (including Arp 55, Arp 299, NGC 5331, and NGC 5257/8) and
thus we are left with four starburst-dominated objects with good size
estimates. They are Arp 193, VV 114, NGC 1614, and IRAS 10565.

In total we have 21 LIRGs with median \LTIR\ of $10^{11.5}$ \Lsun\ and a
median physical diameter of 0.8 kpc. This sample is dominated by the
12 galaxies with Pa$\alpha$ images, and these galaxies by themselves
should provide an unbiased estimate of sizes of the star forming
regions in LIRGs. The estimated sizes of the galaxies measured in the
radio and in CO are consistent with those from Pa$\alpha$.

\subsection{ULIRGs}\label{sec:data_local_ULIRGs}
Despite the many local ULIRGs, selecting a sample to study the
physical sizes of the starburst regions is complicated by two
issues. First, a majority of these ULIRGs harbor AGN and thus do not
represent a starburst environment. Second, the sizes of the
IR-emitting regions are significantly smaller than the optical and
near-IR extents of these galaxies and the existing mid-IR data in the
literature (e.g. SINGS) do not resolve the ULIRGs' IR emission. Our
compilation therefore relies on radio and submillimeter
interferometric observations of four ULIRGs that are known to have
their IR luminosity dominated by star-forming emission: VII Zw 31, IRAS
F23365+3604, Arp 220, and IRAS F17207-0014. Their IR luminosities
according to \citet{Sanders03} are in the range of
$10^{12.00}-10^{12.45}$ \Lsun.  

Arp 220 has two nuclei $\sim$370 pc apart; the western
nucleus is slightly more luminous than the eastern nucleus
\citep{Rovilos03}. This galaxy was studied with Very Long Baseline
Interferometry (VLBI) at 18 cm by \citet{Smith98} who found the
western nucleus to contain most of the individual 18 cm sources (13
sources in the western nucleus vs. three in the eastern nucleus). They
report that the sources in the western nucleus are confined within 75
pc $\times 150$ pc. Circularizing these values gives a diameter of
$0.1$ kpc for the western nucleus alone. The 1.6 
and 5 GHz MERLIN radio maps by \citet{Rovilos03} indicate both nuclei
have similar physical sizes. We estimate the physical
size of Arp 220 from the 5 GHz map where the two nuclei are clearly
separated. Combining the emission regions from both nuclei, we adopt an
effective diameter of 0.2 kpc for the starburst region of Arp 220.

The sizes of VII Zw 31 and  IRAS F23365+3604 are reported by \citet{DS98}
based on CO ($1-0$) observations; the IRAS F17207-0014 size by
\citet{Iono09} is based on CO ($3-2$) observations. Again, we treat these
CO-derived sizes as upper limits for the size of the IR-emitting extent
of the galaxy. 

\clearpage
\begin{center}
\begin{deluxetable}{cccccc}
\tablewidth{0pt}
\tablecaption{Local Compilation of Star-Forming Galaxies}
\tablehead{Source & IRAS ID & Distance\tablenotemark{a} & \LTIR\tablenotemark{a} & Diameter & References\tablenotemark{b} \\
 & & (Mpc) & (logL$_{\odot}$) & (kpc)}
\startdata
\sidehead{\textbf{Normal SF Galaxies}} 
NGC 2976	& F09431+6809 & 3.8	& 8.61	& 1.5	& 1\\
NGC 4826 	& F12542+2157 & 6.0 	& 9.14 	& 1.2 	& 1\\
NGC 2403 	& F07320+6543 & 3.8 	& 9.25 	& 3.0	& 1\\
NGC 925 	& F02242+3321 & 9.8 	& 9.46 	& 6.1 	& 1\\
NGC 1512 	& ... 	      & 11.3 	& 9.49 	& 3.2 	& 1\\
NGC 5866 	& F15051+5557 & 13.0 	& 9.51 	& 1.8 	& 1\\
NGC 2841 	& ... 	      & 10.5 	& 9.52 	& 4.9 	& 1\\
NGC 4559 	& F12334+2814 & 11.9 	& 9.62 	& 5.0 	& 1\\
NGC 4736 	& 12485+4123  & 5.7 	& 9.79 	& 1.7 	& 1\\
NGC 3198 	& F10168+4547 & 14.7 	& 9.81 	& 3.8 	& 1\\
NGC 3184 	& 10152+4140  & 11.9 	& 9.86 	& 6.5 	& 1\\
NGC 3351 	& F10413+1157 & 10.8 	& 9.89 	& 2.2 	& 1\\
NGC 3938 	& F11502+4423 & 13.1 	& 9.99 	& 4.7 	& 1\\
NGC 4569 	& F12343+1326 & 17.8 	& 10.08 & 3.1 	& 1\\
NGC 5055 	& F13135+4217 & 8.4 	& 10.15 & 4.3 	& 1\\
NGC 5033 	& F13111+3651 & 14.7 	& 10.19 & 1.1 	& 1\\
NGC 3627 	& F11176+1315 & 9.3 	& 10.44 & 3.5 	& 1\\
NGC 5194 	& F13277+4727 & 8.8 	& 10.48 & 5.7 	& 1\\
NGC 7331 	& F22347+3409 & 16.2 	& 10.64 & 4.7 	& 1\\
\sidehead{\textbf{LIRGs}} 
NGC 23 		& F00073+2538 & 63.9 	& 11.11 & 1.2 	& 2\\
NGC 6701 	& F18425+6036 & 60.6 	& 11.11 & 0.7 	& 2\\
UGC 1845 	& F02208+4744 & 66.4 	& 11.13 & 0.81 	& 2\\
NGC 5936 	& F15276+1309 & 65.1 	& 11.13 & 0.56 	& 2\\
MCG +02-20-003 	& F07329+1149 & 72.4 	& 11.14 & 0.78 	& 2\\
NGC 2369 	& F07160-6215 & 47.1 	& 11.16 & 0.82 	& 2\\
ESO 320-G030 	& F11506-3851 & 40.4 	& 11.16 & 0.91 	& 2\\
IC 5179 	& F22132-3705 & 50.0 	& 11.22 & 1.6 	& 2\\
NGC 2388 	& F07256+3355 & 61.9 	& 11.29 & 0.84 	& 2\\
NGC 7771 	& F23488+1949 & 61.2 	& 11.40	& 1.04 	& 2\\
MCG +12-02-001 	& F00506+7248 & 68.9 	& 11.50	& 0.8 	& 2\\
...  		& F03359+1523 & 146.9 	& 11.53 & 0.07 	& 3\\
NGC 1614 	& F04315-0840 & 67.1 	& 11.66 & 0.9 	& 4\\
UGC 2369     	& F02512+1446 & 130.7 	& 11.66 & 0.08 	& 3\\
Arp 236 	& F01053-1746 & 84.2 	& 11.71 & 2.6 	& 4\\
Arp 193 	& F13182+3424 & 107.1 	& 11.73 & 0.8 	& 4\\
UGC 4881     	& F09126+4432 & 172.7 	& 11.75 & 0.14 	& 3\\
Arp 299 	& F11257+5850 & 51.2 	& 11.94 & 0.42 	& 4\\
...  		& F17132+5313 & 218.9 	& 11.95 & 0.14 	& 3\\
...             & F15163+4255 & 180.8   & 11.95 & 0.22  & 3\\
... 		& F10565+2448 & 188.9 	& 11.99 & 0.8 	& 4\\
\sidehead{\textbf{ULIRGs}} 
VII Zw 31 	& F05081+7936 & 230.2 	& 12.00	& 2.1 	& 5\\
...     	& F23365+3604 & 269.8 	& 12.19 & 1.1   & 5\\
Arp 220 	& F15327+2340 & 85.6 	& 12.27 & 0.2 	& 6\\
...  		& F17207-0014 & 188.2 & 12.45 & $<0.9$ 	& 4\\
\enddata
\tablenotetext{a}{Distance and \LTIR\ from \citet{Sanders03} and
  adjust to match our cosmology.}
\tablenotetext{b}{Diameters references 1. measured from {\it Spitzer}
  MIPS 24 $\micron$ imaging taken by SINGS
  \citep[e.g.,]{Kennicutt03, Calzetti07}, 2. measured from {\it
    Hubble} NICMOS Pa-$\alpha$ taken by \citet{AH06}, 3. 8.4 GHz
  radio sizes given by \citet{Condon91}, 4. CO ($3-2$) sizes given by
  \citet{Iono09}, 5. CO ($2-1$) or CO ($1-0$) sizes given by
  \citet{DS98}, 6. 5 GHz radio size based on \citet{Rovilos03}.}
\label{table_local}
\end{deluxetable}
\end{center}

\clearpage
\begin{center}
\begin{deluxetable}{cccccccc}
\tablewidth{0pt}
\tablecaption{High-Redshift Compilation of Star-Forming Galaxies}
\tablehead{R.A. & Decl. & Source & z & S$_{24~\micron}$ & S$_{1.4~{\rm
      GHz}}$ & Diameter & References\tablenotemark{a} \\ 
 (J2000) & (J2000) & & & ($\mu$Jy) & ($\mu$Jy) & (kpc)}
\startdata
10 51 46.61 & 57 20 33.4 & RGJ105146.61+572033.4  & 2.383 & $298 \pm 16$  & $33.5 \pm 5.8$    &  4.2 & 7\\
10 51 51.69 & 57 26 36.1 & SMMJ105151.69+572636.1 & 1.147 & $314 \pm 24$  & $106 \pm 6$       &  6.1 & 8, 11\\
10 51 54.19 & 57 24 14.6 & RGJ105154.19+572414.6  & 0.922 & $510 \pm 22$  & $45.4 \pm 6.3$    &  4.0 & 7\\
10 51 55.47 & 57 23 12.8 & SMMJ105155.47+572312.8 & 2.686 & $104 \pm 14$  & $51.0 \pm 4.3$      &  2.2 & 8, 11\\
10 51 58.02 & 57 18 00.3 & SMMJ105158.02+571800.3 & 2.239 & $303 \pm 32$  & $92.3 \pm 4.5$    &  6.7 & 8, 11\\
10 51 59.90 & 57 18 02.4 & RGJ105159.90+571802.4  & 1.047 & $738 \pm 27$  & $74.5 \pm 5.6$    &  5.4 & 7\\
10 52 01.25 & 57 24 45.8 & SMMJ105201.25+572445.8 & 2.148 & $217 \pm 16$  & $78.9 \pm 4.7$    &  3.3 & 8, 11\\
12 35 49.44 & 62 15 36.8 & SMM123549.44+621536.8  & 2.203 & 630           & $74.6 \pm 9.5$    &  2.5 & 2, 3, 6\\
12 36 07.13 & 62 13 28.6 & J123607+621328         & 0.435 & $ 259 \pm  6$ & $80.1 \pm 6.0$ &  5.3 & 10, 12, 14\\
12 36 15.60 & 62 09 46.4 & J123615+620946         & 1.263 & $ 101 \pm  4$ & $55.3 \pm 9.7$ &  5.2 & 10, 12, 14\\
12 36 16.15 & 62 15 13.7 & SMMJ123616.15+621513.7 & 2.578 & $313 \pm 7$   & $53.9 \pm 8.4$    &  8.0 & 2, 10, 12, 14\\
12 36 17.07 & 62 10 11.2 & J123617+621011         & 0.845 & $  88 \pm  6$ & $65.3 \pm 8.3$ &  3.6 & 10, 12, 14\\
12 36 18.32 & 62 15 50.5 & J123618+621550         & 1.870 & $ 330 \pm  7$ & $164.4 \pm 6.9$ &  2.6 & 10, 12, 14\\
12 36 19.46 & 62 12 52.6 & J123619+621252         & 0.473 & $ 976 \pm 12$ & $65.3 \pm 4.8$ &  1.8 & 10, 12, 14\\
12 36 22.65 & 62 16 29.7 & SMMJ123622.65+621629.7 & 2.466 & $414 \pm 7$   & $70.9 \pm 8.7$    &  9.7 & 2, 10, 12, 14\\
12 36 26.52 & 62 08 35.4 & BzK-4171               & 1.465 & 142           & 34                & 11.3 & 9, 13\\
12 36 29.13 & 62 10 45.8 & SMMJ123629.13+621045.8 & 1.013 & $724 \pm 12$  & $81.4 \pm 8.7$    &  6.6 & 1, 2, 10\\
12 36 30.02 & 62 09 23.7 & J123630+620923         & 0.953 & $ 223 \pm  6$ & $41.1 \pm   4.7$ &  3.7 & 10, 12, 14\\
12 36 30.12 & 62 14 28.0 & BzK-16000              & 1.522 & 183           & 19                & 10.9 & 9, 13\\
12 36 33.67 & 62 10 05.8 & J123633+621005         & 1.016 & $ 581 \pm  9$ & $58.5 \pm   9.1$ &  7.5 & 10, 12, 14\\
12 36 34.45 & 62 12 12.9 & J123634+621213         & 0.456 & $1290 \pm  8$ & $224.7 \pm  10.9$ &  5.4 & 10, 12, 14\\
12 36 34.49 & 62 12 41.0 & J123634+621241         & 1.219 & $ 446 \pm  5$ & $201.1 \pm  10.3$ &  6.4 & 10, 12, 14\\
12 36 35.57 & 62 14 24.0 & J123635+621424         & 2.011 & $1480 \pm 10$ & $76.0 \pm   7.9$ &  1.9 & 10, 12, 14\\
12 36 41.52 & 62 09 48.2 & J123641+620948         & 0.518 & $ 433 \pm  6$ & $83.5 \pm   8.1$ &  2.9 & 10, 12, 14\\
12 36 45.89 & 62 07 54.1 & RGJ123645.89+620754.1  & 1.433 & $369 \pm 7$   & $83.4 \pm 9.8$    &  4.2 & 7, 10\\
12 36 46.64 & 62 08 33.3 & J123646+620833         & 0.971 & $ 982 \pm  8$ & $81.7 \pm   5.1$ &  4.9 & 10, 12, 14\\
12 36 49.68 & 62 13 12.9 & J123649+621313         & 0.475 & $ 371 \pm 10$ & $59.3 \pm   8.2$ &  4.6 & 10, 12, 14\\
12 36 50.01 & 62 08 01.6 & J123650+620801         & 0.559 & $ 223 \pm  6$ & $34.1 \pm   8.1$ &  3.5 & 10, 12, 14\\
12 36 51.11 & 62 10 30.8 & J123651+621030         & 0.410 & $ 984 \pm  9$ & $86.8 \pm   8.2$ &  5.1 & 10, 12, 14\\
12 36 53.37 & 62 11 39.6 & RGJ123653.37+621139.6  & 1.275 & $322 \pm 6$   & $86.7 \pm 8.3$    &  5.8 & 7, 10\\
12 36 55.73 & 62 09 17.4 & J123655+620917         & 0.419 & $ 846 \pm  9$ & $78.3 \pm   8.8$ &  2.6 & 10, 12, 14\\
12 36 55.93 & 62 08 08.2 & J123655+620808         & 0.792 & $ 832 \pm  7$ & $118.0 \pm   5.6$ &  4.1 & 10, 12, 14\\
12 36 59.92 & 62 14 50.1 & J123659+621449         & 0.761 & $ 466 \pm  5$ & $42.1 \pm   9.2$ &  5.7 & 10, 12, 14\\
12 37 04.08 & 62 07 55.3 & J123704+620755         & 1.253 & $ 497 \pm  9$ & $63.1 \pm  10.2$ &  6.5 & 10, 12, 14\\
12 37 05.88 & 62 11 53.7 & J123705+621153         & 0.902 & $ 655 \pm  8$ & $49.4 \pm   8.1$ &  9.1 & 10, 12, 14\\
12 37 07.21 & 62 14 08.1 & SMMJ123707.21+621408.1 & 2.484 & $235 \pm 8$   & $45.0 \pm 7.9$   &  7.3 & 2, 3, 10\\
12 37 08.32 & 62 10 56.0 & J123708+621056         & 0.422 & $ 648 \pm  7$ & $66.8 \pm   8.1$ &  3.0 & 10, 12, 14\\
12 37 10.60 & 62 22 34.6 & RGJ123710.60+622234.6  & 1.522 & $227 \pm 39$  & $38.3 \pm 10.1$   &  3.4 & 7\\
12 37 11.98 & 62 13 25.7 & SMMJ123711.98+621325.7 & 1.992 & $225 \pm 7$   & $53.9 \pm 8.1$    &  6.9 & 1, 2, 10\\
12 37 13.58 & 62 16 03.7 & J123713+621603         & 0.938 & $ 208 \pm  6$ & $61.2 \pm   9.1$ &  5.5 & 10, 12, 14\\
12 37 14.34 & 62 15 58.8 & J123714+621558         & 0.567 & $ 155 \pm  5$ & 51               &  7.6 & 10, 12\\
12 37 16.58 & 62 16 43.2 & J123716+621643         & 0.557 & $ 512 \pm  6$ & $79.1 \pm   5.2$ &  4.5 & 10, 12, 14\\
12 37 16.81 & 62 10 07.3 & J123716+621007         & 0.411 & $ 583 \pm  8$ & $92.7 \pm   9.3$ &  3.0 & 10, 12, 14\\
12 37 21.45 & 62 13 46.1 & J123721+621346         & 1.019 & $ 217 \pm  6$ & $41.6 \pm   8.7$ &  4.4 & 10, 12, 14\\
12 37 51.82 & 62 15 20.2 & BzK-17999              & 1.414 & 230           & 34                &  6.4 & 9, 13\\
14 18 03.55 & 52 30 22.3 & EGS 12007881           & 1.161 & 327.6         & ...               &   16 & 4, 13\\
14 20 05.43 & 53 01 15.5 & EGS 13035123           & 1.115 & 571.3         & ...               &   13 & 4, 13\\
16 36 58.19 & 41 05 23.8 & SMMJ163658.19+410523.8 & 2.454 & $330 \pm 55$  & $92 \pm 16$       &  3.2 & 2, 3, 5\\
\enddata
\tablenotetext{a}{1. \citet{Chapman04}, 2. \citet{Chapman05},
  3. \citet{Tacconi06}, 4. \citet{Tacconi10}, 5. \citet{Hainline09},
  6. \citet{MD09}, 7. \citet{Casey09}, 8. \citet{BiggsIvison08},
  9. \citet{Daddi10}, 10. GOODS {\it Spitzer} Legacy Data, Dickinson
  et al., in prep., 11. \citet{Ivison07}, 12. \citet{Muxlow05},
  13. Weiner, private comm., 14. \citet{Morrison10}}
\label{table_highz}
\end{deluxetable}
\end{center}


\clearpage
\begin{thebibliography}{}

\bibitem[Agertz et al.(2009)]{Agertz09} Agertz, O., Teyssier, 
R., \& Moore, B.\ 2009, \mnras, 397, L64 

\bibitem[Alonso-Herrero et al.(2006)]{AH06} Alonso-Herrero, 
A., Rieke, G.~H., Rieke, M.~J., Colina, L., P{\'e}rez-Gonz{\'a}lez, P.~G., 
\& Ryder, S.~D.\ 2006, \apj, 650, 835 

\bibitem[Biggs 
\& Ivison(2008)]{BiggsIvison08} Biggs, A.~D., \& Ivison, R.~J.\ 2008,
  \mnras, 385, 893 

\bibitem[Bothwell et al.(2010)]{Bothwell09} Bothwell, M.~S., et 
al.\ 2010, \mnras, 525 

\bibitem[Blain et al.(2002)]{Blain02} Blain, A.~W., Smail, I., 
Ivison, R.~J., Kneib, J.-P., \& Frayer, D.~T.\ 2002, \physrep, 369, 111 

\bibitem[Brandl et al.(2006)]{Brandl06} Brandl, B.~R., et al.\ 
2006, \apj, 653, 1129 

\bibitem[Calzetti et al.(2007)]{Calzetti07} Calzetti, D., et al.\ 
2007, \apj, 666, 870 

\bibitem[Calzetti et al.(2010)]{Calzetti10} Calzetti, D., et al.\ 
2010, \apj, 714, 1256 

\bibitem[Casey et al.(2009)]{Casey09} Casey, C.~M., et al.\ 
2009, \mnras, 399, 121 

\bibitem[Carilli et al.(2010)]{Carilli10} Carilli, C.~L., et al.\ 
2010, \apj, 714, 1407 

\bibitem[Chapman et al.(2004)]{Chapman04} Chapman, S.~C., Smail, 
I., Windhorst, R., Muxlow, T., \& Ivison, R.~J.\ 2004, \apj, 611, 732 

\bibitem[Chapman et al.(2005)]{Chapman05} Chapman, S.~C., Blain, 
A.~W., Smail, I., \& Ivison, R.~J.\ 2005, \apj, 622, 772 

\bibitem[Condon et al.(1991)]{Condon91} Condon, J.~J., Huang, 
Z.-P., Yin, Q.~F., \& Thuan, T.~X.\ 1991, \apj, 378, 65 

\bibitem[Condon et al.(1998)]{Condon98} Condon, J.~J., Cotton, 
W.~D., Greisen, E.~W., Yin, Q.~F., Perley, R.~A., Taylor, G.~B., 
\& Broderick, J.~J.\ 1998, \aj, 115, 1693 

\bibitem[Coppin et al.(2006)]{Coppin06} Coppin, K., et al.\ 
2006, \mnras, 372, 1621 

\bibitem[Daddi et al.(2010)]{Daddi10} Daddi, E., et al.\ 2010, 
\apj, 713, 686 

\bibitem[Dale et al.(2006)]{Dale06} Dale, D.~A., et al.\ 2006, 
\apj, 646, 161 

\bibitem[Dale et al.(2007)]{Dale07} Dale, D.~A., et al.\ 2007, 
\apj, 655, 863 

\bibitem[Dasyra et al.(2009)]{Dasyra09} Dasyra, K.~M., et al.\ 
2009, \apj, 701, 1123 

\bibitem[Dav{\'e} et al.(2010)]{Dave10} Dav{\'e}, R., 
Finlator, K., Oppenheimer, B.~D., Fardal, M., Katz, N., Kere{\v s}, D., 
\& Weinberg, D.~H.\ 2010, \mnras, 404, 1355 

\bibitem[Downes 
\& Solomon(1998)]{DS98} Downes, D., \& Solomon, P.~M.\ 1998, \apj, 507, 615 

\bibitem[Engelbracht et al.(2008)]{Engelbracht08} Engelbracht, C.~W., 
Rieke, G.~H., Gordon, K.~D., Smith, J.-D.~T., Werner, M.~W., Moustakas, J., 
Willmer, C.~N.~A., \& Vanzi, L.\ 2008, \apj, 678, 804 

\bibitem[Erb et al.(2006)]{Erb06} Erb, D.~K., Shapley, A.~E., 
Pettini, M., Steidel, C.~C., Reddy, N.~A., 
\& Adelberger, K.~L.\ 2006, \apj, 644, 813 

\bibitem[Farrah et al.(2008)]{Farrah08} Farrah, D., et al.\ 
2008, \apj, 677, 957 

\bibitem[Gao 
\& Solomon(2004)]{GS04} Gao, Y., \& Solomon, P.~M.\ 2004, \apj, 606, 271 

\bibitem[Giavalisco et al.(2004)]{Giavalisco04} Giavalisco, M., et 
al.\ 2004, \apjl, 600, L93 

\bibitem[Genzel et al.(2010)]{Genzel10} Genzel, R., et al.\ 
2010, arXiv:1003.5180 

\bibitem[Greve et al.(2003)]{Greve03} Greve, T.~R., Ivison, 
R.~J., \& Papadopoulos, P.~P.\ 2003, \apj, 599, 839 

\bibitem[Greve et al.(2004)]{Greve04} Greve, T.~R., Ivison, 
R.~J., Bertoldi, F., Stevens, J.~A., Dunlop, J.~S., Lutz, D., 
\& Carilli, C.~L.\ 2004, \mnras, 354, 779 

\bibitem[Hailey-Dunsheath et al.(2010)]{HD10} 
Hailey-Dunsheath, S., Nikola, T., Stacey, G.~J., Oberst, T.~E., Parshley, 
S.~C., Benford, D.~J., Staguhn, J.~G., 
\& Tucker, C.~E.\ 2010, \apjl, 714, L162 

\bibitem[Hainline et al.(2009)]{Hainline09} Hainline, L.~J., 
Blain, A.~W., Smail, I., Frayer, D.~T., Chapman, S.~C., Ivison, R.~J., 
\& Alexander, D.~M.\ 2009, \apj, 699, 1610 

\bibitem[Iono et al.(2009)]{Iono09} Iono, D., et al.\ 2009, 
\apj, 695, 1537 

\bibitem[Ivison et al.(2007)]{Ivison07} Ivison, R.~J., et al.\ 
2007, \mnras, 380, 199 

\bibitem[Ivison et al.(2010a)]{Ivison10a} Ivison, R.~J., et al.\ 
2010a, \mnras, 402, 245 

\bibitem[Ivison et al.(2010b)]{Ivison10b} Ivison, R.~J., et al.\ 
2010b, arXiv:1005.1072 

\bibitem[Ivison et al.(2010c)]{Ivison10c} Ivison, R.~J., et al.\ 
2010c, arXiv:1005.1071 

\bibitem[Juneau et al.(2009)]{Juneau09} Juneau, S., Narayanan, 
D.~T., Moustakas, J., Shirley, Y.~L., Bussmann, R.~S., Kennicutt, R.~C., 
\& Vanden Bout, P.~A.\ 2009, \apj, 707, 1217 

\bibitem[Kennicutt(1998)]{Kennicutt98} Kennicutt, R.~C., Jr.\ 1998, 
\apj, 498, 541 

\bibitem[Kennicutt et al.(2003)]{Kennicutt03} Kennicutt, R.~C., 
Jr., et al.\ 2003, \pasp, 115, 928 

\bibitem[Lagache et 
al.(2005)]{Lagache05} Lagache, G., Puget, J.-L., \& Dole, H.\ 2005, \araa, 43, 727 

\bibitem[Le Floc'h et al.(2005)]{LeFloch05} Le Floc'h, E.,
et al.\ 2005, \apj, 632, 169

\bibitem[Lehnert et al.(2009)]{Lehnert09} Lehnert, M.~D., 
Nesvadba, N.~P.~H., Tiran, L.~L., Matteo, P.~D., van Driel, W., Douglas, 
L.~S., Chemin, L., \& Bournaud, F.\ 2009, \apj, 699, 1660 

\bibitem[Men{\'e}ndez-Delmestre et al.(2009)]{MD09} 
Men{\'e}ndez-Delmestre, K., et al.\ 2009, \apj, 699, 667 

\bibitem[Momjian et al.(2006)]{Momjian06} Momjian, E., Romney, 
J.~D., Carilli, C.~L., \& Troland, T.~H.\ 2006, \apj, 653, 1172 

\bibitem[Morrison et al.(2010)]{Morrison10} Morrison, G.~E., Owen, 
F.~N., Dickinson, M., Ivison, R.~J., \& Ibar, E.\ 2010, \apjs, 188, 178 

\bibitem[Murphy et al.(2006)]{Murphy06} Murphy, E.~J., et al.\ 
2006, \apj, 638, 157 

\bibitem[Muxlow et al.(2005)]{Muxlow05} Muxlow, T.~W.~B., et 
al.\ 2005, \mnras, 358, 1159 

\bibitem[Muzzin et al.(2010)]{Muzzin10} Muzzin, A., van Dokkum, 
P., Kriek, M., Labbe, I., Cury, I., Marchesini, D., 
\& Franx, M.\ 2010, arXiv:1003.3479 

\bibitem[Neri et al.(2003)]{Neri03} Neri, R., et al.\ 2003, 
\apjl, 597, L113 

\bibitem[Papovich et al.(2007)]{Papovich07} Papovich, C., et al.\ 
2007, \apj, 668, 45 

\bibitem[P{\'e}rez-Gonz{\'a}lez et al.(2005)]{PPG05} 
P{\'e}rez-Gonz{\'a}lez, P.~G., et al.\ 2005, \apj, 630, 82

\bibitem[Pope et al.(2006)]{Pope06} Pope, A., et al.\ 2006, 
\mnras, 370, 1185 

\bibitem[Reddy et al.(2010)]{Reddy10} Reddy, N.~A., Erb, D.~K., 
Pettini, M., Steidel, C.~C., \& Shapley, A.~E.\ 2010, \apj, 712, 1070 

\bibitem[Rieke et al.(2009)]{Rieke09} Rieke, G.~H., 
Alonso-Herrero, A., Weiner, B.~J., P{\'e}rez-Gonz{\'a}lez, P.~G., Blaylock, 
M., Donley, J.~L., \& Marcillac, D.\ 2009, \apj, 692, 556 

\bibitem[Rigby et al.(2008)]{Rigby08} Rigby, J.~R., et al.\ 
2008, \apj, 675, 262 

\bibitem[Rodighiero et 
al.(2010)]{Rodighiero10} Rodighiero, G., et al.\ 2010, \aap, 515, A8 

\bibitem[Rovilos et al.(2003)]{Rovilos03} Rovilos, E., Diamond, 
P.~J., Lonsdale, C.~J., Lonsdale, C.~J., 
\& Smith, H.~E.\ 2003, \mnras, 342, 373 

\bibitem[Sanders et al.(2003)]{Sanders03} Sanders, D.~B., 
Mazzarella, J.~M., Kim, D.-C., Surace, J.~A., 
\& Soifer, B.~T.\ 2003, \aj, 126, 1607 

\bibitem[Sargent et al.(2010)]{Sargent10} Sargent, M.~T., et al.\ 
2010, \apjl, 714, L190 

\bibitem[Schmidt(1959)]{Schmidt59} Schmidt, M.\ 1959, \apj, 129, 
243 

\bibitem[Scott et al.(2002)]{Scott02} Scott, S.~E., et al.\ 
2002, \mnras, 331, 817 

\bibitem[Smith et al.(1998)]{Smith98} Smith, H.~E., Lonsdale, 
C.~J., Lonsdale, C.~J., \& Diamond, P.~J.\ 1998, \apjl, 493, L17 

\bibitem[Swinbank et al.(2010a)]{Swinbank10a} Swinbank, M., et al.\ 
2010a, arXiv:1002.2518 

\bibitem[Swinbank et al.(2010b)]{Swinbank10b} Swinbank, A.~M., et 
al.\ 2010b, \nat, 464, 733 

\bibitem[Symeonidis et al.(2009)]{Symeonidis09} Symeonidis, M., 
Page, M.~J., Seymour, N., Dwelly, T., Coppin, K., McHardy, I., Rieke, 
G.~H., \& Huynh, M.\ 2009, \mnras, 397, 1728 

\bibitem[Tacconi et al.(2006)]{Tacconi06} Tacconi, L.~J., et al.\ 
2006, \apj, 640, 228 

\bibitem[Tacconi et al.(2008)]{Tacconi08} Tacconi, L.~J., et al.\ 
2008, \apj, 680, 246 

\bibitem[Tacconi et al.(2010)]{Tacconi10} Tacconi, L.~J., et al.\ 
2010, \nat, 463, 781 

\bibitem[Takagi et al.(2010)]{Takagi10} Takagi, T., et al.\ 2010,
  \aap, 514, A5  

\bibitem[Thrall et al.(2007)]{Thrall07} Thrall, H., Muxlow, 
T.~W.~B., Beswick, R.~J., 
\& Richards, A.~M.~S.\ 2007, Deepest Astronomical Surveys, 380, 291 

\bibitem[Wilson et al.(2008)]{Wilson08} Wilson, C.~D., et al.\ 
2008, \apjs, 178, 189 

\bibitem[Younger et al.(2010)]{Younger10} Younger, J.~D., et al.\ 
2010, \mnras, 984 

\end{thebibliography}
\end{document}